\documentclass[aps,prl,twocolumn,showpacs,superscriptaddress,groupedaddress]{revtex4}  
\usepackage{graphicx}  
\usepackage{dcolumn}  
\usepackage{bm}       
\usepackage{amssymb} 
\usepackage{multirow}
\usepackage{epsfig}
\usepackage{epstopdf}
\usepackage{amsmath,booktabs,array}
\usepackage[hyperindex,breaklinks]{hyperref} 
\usepackage{atlasphysics}
\usepackage{subfigure}

\newcommand{\metpaul}{\mbox{$\rlap{\kern0.15em/}E_T$}}
\newcommand{\WW}{$WW$}

\newcommand{\pythia}{{\sc pythia}}

\newcommand {\mtww}{\ensuremath{m_{\rm T}^{WW}}}
\newcommand{\ww}{\ensuremath{WW \rightarrow \ell \nu \ell^{\prime} \nu^{\prime}}}
\def\KKG{\ensuremath{G^*_{\rm bulk}}}
\def\G{\ensuremath{G^*}}

\def\elel{\ensuremath{ee}}
\def\elmu{\ensuremath{e\mu}}
\def\mum{\ensuremath{\mu\mu}}

\usepackage{preprintcover} 
\PreprintCoverPaperTitle{Search for new phenomena in the $WW \to l\nu l'\nu'$ final state in $pp$ collisions at $\sqrt{s} = 7$ \TeV\ with the ATLAS detector} 
\PreprintIdNumber{CERN-PH-EP-2012-197}  
\PreprintCoverAbstract{This Letter reports a search for a heavy particle that decays to $WW$  using events produced in $pp$ collisions at 
$\sqrt{s}=7\TeV$. The data were recorded in 2011 by the ATLAS detector  
and correspond to an integrated luminosity of 4.7~\ifb. 
\ww~($\ell, \ell^{\prime} = e$ or $\mu$)  final states are considered
and the distribution of the transverse mass of the $WW$ candidates is
found to be consistent with Standard Model expectations. 
Upper limits on the production cross section times branching ratio into $W$ boson pairs are set for Randall-Sundrum and bulk Randall-Sundrum gravitons,
which result in observed 95\% CL lower limits on the masses of the
two particles of 1.23 \TeV\ and 0.84 \TeV, respectively.}
\PreprintJournalName{PLB} 

\begin{document}

\title{Search for new phenomena in the $WW \to l\nu l'\nu'$ final state in $pp$ collisions at $\sqrt{s} = 7$ \TeV\ with the ATLAS detector}
\author{The ATLAS Collaboration}
\date{\today}
\begin{abstract}
This Letter reports a search for a heavy particle that decays to $WW$  using events produced in $pp$ collisions at 
$\sqrt{s}=7\TeV$. The data were recorded in 2011 by the ATLAS detector  
and correspond to an integrated luminosity of 4.7~\ifb. 
\ww~($\ell, \ell^{\prime} = e$ or $\mu$)  final states are considered
and the distribution of the transverse mass of the $WW$ candidates is
found to be consistent with Standard Model expectations. 
Upper limits on the production cross section times branching ratio into $W$ boson pairs are set for Randall-Sundrum and bulk Randall-Sundrum gravitons,
which result in observed 95\% CL lower limits on the masses of the
two particles of 1.23 \TeV\ and 0.84 \TeV, respectively.
\end{abstract}

\maketitle

The existence of new phenomena can be probed by studying heavy gauge boson pair production.  
Heavy particles that can decay to gauge boson pairs are predicted in many scenarios of physics beyond the Standard 
Model (SM), including the Extended Gauge Model (EGM)~\cite{egm}, Extra 
Dimensions~\cite{extraDimension1, extraDimension2, extraDimension3, extraDimension4, extraDimension5}, 
and Technicolor models~\cite{TC1, TC2, TC3}. 
This paper describes a search for resonant $WW$ production 
in the \ww~($\ell, \ell' = e$ or $\mu$) decay channel using a data 
sample corresponding to an integrated luminosity of 4.7 \ifb, collected by the ATLAS detector during 2011 at a center-of-mass energy of 7 \TeV. 
A spin-2 Randall-Sundrum (RS) graviton 
model~\cite{extraDimension1} and one of its extensions, the bulk RS graviton model~\cite{brs_model}, are used as benchmarks to interpret the analysis result. 

The original RS model (RS1) was proposed to solve the hierarchy problem. 
It postulates a warped 5-dimensional universe, where the SM particles are localized on the \TeV\ brane and the graviton is located on the Planck brane.  
In this model gravitons can propagate in the extra dimension, 
leading to a Kaluza-Klein tower of states which can be detected as massive spin-2 resonances that couple to all SM particles. 
The resonance with the lowest mass is known as the RS graviton \G. 
The model has two parameters: the graviton mass $m_{\G}$, and the dimensionless coupling $\kappa/\tilde{M}_{\rm pl}$, 
where $\kappa$ is the curvature of the warped fifth dimension and $\tilde{M}_{\rm pl} = M_{\rm pl}/\sqrt{8 \pi}$ is the reduced Planck mass. 

The RS1 model introduces higher-dimensional operators that give excessively large 
contributions to flavour changing neutral current (FCNC) processes 
and to observables related to SM electroweak precision tests. 
An extension of the RS1 model, the bulk RS model, has been proposed to address this issue.
In this model, the SM fields are also allowed to propagate in the extra dimension: 
the first and second generation fermions are chosen to be localized near the Planck brane, 
while the top-quark and the Higgs boson are localized near the \TeV\ brane to account for the large top-quark Yukawa coupling. 
In this scenario, FCNCs and contributions to electroweak observables from higher-dimensional operators are suppressed, 
the graviton (here denoted by \KKG) production and decay via light fermion channels is highly suppressed, the 
probability for the graviton to decay into photons is negligible, and the coupling to heavy particles, such as top-quark, $W$, $Z$ and Higgs bosons is strongly enhanced. In this model the branching ratio of $\KKG \to WW$ is about 15\%.

Direct searches for a heavy \WW~resonance have been performed by the CDF and D0 
collaborations at the Tevatron. The D0 collaboration explored diboson resonant production using the $\ell\nu \ell' \nu'$ 
and $\ell \nu j j$ final states~\cite{D0_WZ}; these searches excluded an RS graviton with a mass between 300 GeV and 754 GeV, 
assuming $\kappa/ \tilde{M}_{\rm pl}=0.1$. 
The CDF collaboration also searched for resonant \WW\ production in the $e \nu j j$ final state, 
resulting in a lower limit of 607 GeV on the mass of an RS graviton~\cite{CDF_WW}, 
assuming the same coupling strength $\kappa/\tilde{M}_{\rm pl}=0.1$. 
No previous work on searches for \KKG\ has been published.

The ATLAS detector~\cite{atlas_detector} is a multi-purpose particle physics detector with forward-backward symmetric cylindrical 
geometry~\cite{geometry}. The inner tracking detector (ID) covers the region $|\eta|<2.5$, and consists of a silicon pixel detector, 
a silicon microstrip detector, and a straw tube tracker with transition radiation detection capability. 
The ID is surrounded by a thin superconducting solenoid providing a $2\:\rm{T}$ axial magnetic field. 
A high-granularity lead/liquid-argon (LAr) sampling calorimeter measures the energy and the position of electromagnetic showers with $\left|\eta\right|<3.2$. 
LAr sampling calorimeters are also used to measure hadronic showers in the end-cap ($1.5<\left|\eta\right|<3.2$) and forward ($3.1<\left|\eta\right|<4.9$) regions, 
while an iron/scintillator tile calorimeter measures hadronic showers in the central region ($\left|\eta\right|<1.7$). 
The muon spectrometer (MS) surrounds the calorimeters and consists of three large superconducting air-core toroids, 
each with eight coils, a system of precision tracking chambers ($\left|\eta\right|<2.7$), and fast tracking chambers for triggering. 
A three-level trigger system selects events to be recorded for offline analysis. 

The data used in this analysis were recorded in 2011 at a centre-of-mass energy of 7 \TeV, selected by 
a single-lepton ($e$ or $\mu$) trigger, with a threshold applied to the 
electron transverse energy, \et, and to the muon transverse momentum, \pt. 
The single-muon trigger required a muon $\pt\ > 18$ \GeV, while for the single-electron trigger the threshold was raised from 20 GeV to 22 GeV for later data. 
The trigger object quality requirements were tightened progressively throughout the data-taking period to cope with the increasing instantaneous luminosity. 
After the application of data-quality requirements, the data set corresponds to a total integrated luminosity of 4.7 \ifb\ with an uncertainty of 3.9\%~\cite{Aad:2011dr,ATLAS-CONF-2011-116}.

The search for resonant $WW$ production is performed in the fully leptonic decay channel. Events are required to contain two oppositely-charged leptons (either electrons or muons) and large missing transverse momentum \met\ due to the presence of neutrinos in the final state. Henceforth this final state is denoted by $\ell \ell' + \met$. 

Events originating from $pp$ collisions are selected by requiring a reconstructed primary interaction vertex with at least three tracks with $\pt\ > 0.4$ \GeV. 
Electron candidates are selected from clustered energy deposits in
the electromagnetic calorimeter with $\et > 25$ GeV and within the ID
fiducial region $|\eta|
< 2.47$, excluding the transition region between barrel and endcap
calorimeters $1.37 < |\eta| < 1.52$. 
A set of electron identification criteria based on the calorimeter shower shape, track quality
and track-matching with the calorimeter cluster, referred to as {\it
  tight}~\cite{electron_performance}, is applied. 
Muon candidates must be reconstructed in both the ID and the MS, and have $\pt\ > 25$ GeV and $|\eta| < 2.4$. 
A minimum number of silicon strip and pixel hits associated to the ID muon track is also required. To ensure good reconstruction quality 
even for very high-\pt\ muons, the charge-to-momentum ratio of the muon tracks reconstructed in the ID and MS 
have to be compatible within five standard deviations. 
Both electron and muon candidates are required to be isolated: the transverse energy deposited in the 
calorimeter in a $\Delta R = \sqrt{(\Delta \eta)^2 + (\Delta \phi)^2} = 0.3$ cone around the lepton track, 
excluding the energy associated to the lepton itself, must be less than 0.14 times the \et\ (\pt) of the electron (muon);
and the scalar sum of the transverse momentum of all tracks with $\pt\ > 1$ GeV reconstructed within $\Delta R = 0.3$ 
around the lepton track, must be less than 0.13 (0.15) times the \et\ (\pt) of the electron (muon). 
Corrections are applied to account for electron energy leakage and energy deposition inside
the isolation cone due to additional $pp$ collisions occurring in the same or neighbouring bunch crossings. 
To ensure the leptons originate from the primary interaction vertex each candidate's longitudinal impact parameter is required to be less than 1 mm, 
and the transverse impact parameter divided by its resolution 
is required to be less than ten for electrons and less than three for muons.  
Any electron reconstructed in a $\Delta R = 0.1$ cone around a muon track is discarded.

Jets tagged as originating from a $b$-quark are used in this analysis to suppress the top background. 
Jets are reconstructed from noise-suppressed three-dimensional topological clusters of calorimeter cells~\cite{jetPerf} using 
the anti-$k_t$ algorithm~\cite{antikt} with radius parameter $R = 0.4$.
Topological clustering extends up to $|\eta| < 4.9$, and clusters are seeded 
by calorimeter cell deposits exceeding the cell noise level by at least four standard deviations. 
Neighbouring cells exceeding the cell noise level by at least two standard deviations are then added to the clusters. 
At least 75\% of the scalar sum of the \pt\ of all the tracks associated to each jet
must belong to tracks associated to the same primary vertex. 

Jet energies are calibrated using $\et$- and $\eta$-dependent correction factors based on Monte Carlo (MC) simulation, 
and validated by collision data studies~\cite{jet_study}. 
Jets are identified as originating from $b$-quarks using an algorithm that combines 
information about the impact parameter significance of tracks in the jet 
with the topology of semi-leptonic $b$- and $c$-hadron decays~\cite{btag}. 
The chosen operating point has an efficiency of 85\% for tagging
$b$-jets in a MC sample of $t\bar{t}$ events, and a mis-tag rate of
less than 5\% for jets from light quarks, $c$-quarks and gluons. A scale factor is
applied to the $b$-tagging efficiency and to the light- and $c$- to $b$-quark
jets mis-tag rate of the MC simulation to reproduce the ones measured
in the data. 
The fiducial kinematic region for well-reconstructed $b$-jets is $\pt
> 20$ GeV and $|\eta| < 2.5$. In order to remove electrons reconstructed as jets, $b$-jet candidates
that lie within a $\Delta R = 0.3$ cone around an electron track are discarded. 

The \met\ is determined by the energy collected by the electromagnetic and hadronic calorimeters, and by 
muon tracks reconstructed in the MS and the ID~\cite{met}. 

Candidate $WW$ events are required to have exactly two oppositely-charged leptons with dilepton invariant mass greater than 
106 GeV to reduce the background contamination from $Z$ boson production. Three different final states are considered based on the lepton flavour, 
namely \elel, \mum, and \elmu. To cope with different background compositions, a different requirement on the \met\ is applied to each final state, which is 
$\met > 30, 60$ and $65$ GeV for \elmu, \elel\ and \mum, respectively. To reject top-quark backgrounds, events with any reconstructed $b$-jets are discarded. 

The SM processes that can mimic the $\ell \ell' +\met$ signature are: electroweak diboson 
pair production, namely $WW$, which is an irreducible background, $WZ/ZZ$ when only two leptons are reconstructed in the final state, and $W\gamma$ when 
the photon is reconstructed as a lepton; top-pair and single-top production, when the $b$-jets in the final state are not identified; 
$W/Z$ production in association with jets, when either one jet is reconstructed as a lepton as for $W+$jets events, or fake \met\ is generated from the mismeasurement of the \pt\ of the leptons or jets; 
and QCD multi-jet production, when two jets are reconstructed as leptons. 
 
The expected background contributions from SM diboson, single-top and \ttbar\ production are estimated using the MC simulation~\cite{atlas_simulation}. 
MC samples are generated at $\sqrt{s} = 7$ \TeV\ using a {\sc geant4}~\cite{geant} simulation of the ATLAS detector. 
To improve the agreement between data and simulation, selection efficiencies are measured
in both data and simulation, and correction factors are applied to the simulation. Furthermore, the simulation is
tuned to reproduce the muon momentum scale and the muon momentum and electron energy resolutions observed in data.  
The MC predictions are normalized to the data sample integrated luminosity, 
except for $W/Z+$jets processes, whose contributions are estimated from data.
$WW$ and \ttbar\ production are simulated using the next-to-leading-order (NLO) generator {\sc mc@nlo} 3.4~\cite{mcnlo}, 
interfaced to {\sc  herwig} 6.510~\cite{herwig} for hadronization and parton showering. 
The gg{\sc2ww}~\cite{gg2ww} program is used to simulate at next-to-next-to-leading order (NNLO) the 
$WW$ production via gluon fusion, which is not implemented in {\sc mc@nlo}; 
{\sc  herwig} 6.510 and {\sc alpgen} 4~\cite{alpgen} are used to simulate at leading order (LO) 
the $WZ/ZZ$ and $W\gamma$ processes respectively, and NLO corrections computed using {\sc mcfm}~\cite{Campbell:2011bn} are then applied; 
$W/Z+$jets processes are simulated at LO using {\sc alpgen} 4 and NNLO corrections computed with {\sc fewz} 2.0~\cite{Gavin:2010az} are applied; 
single-top production is simulated at LO using {\sc acermc}~\cite{acermc}.  

After event selection, top-quark pair production is one of the dominant backgrounds. 
In order to ensure that the MC simulation correctly models the production cross section and kinematics of top-quark events, 
a background dominated control region (denoted by ``top control region'') is defined using the same selection as for the signal region, 
except requiring two reconstructed $b$-jets, instead of zero $b$-jets. 
This region is kinematically close to the signal region, and completely dominated by top-quark pair production. 
The number of observed events in the top control region in data is 322 for $ee$, 370 for $\mu\mu$  and 1303 for $e\mu$ channels, 
to be compared with the MC prediction of $306 \pm 97$, $400 \pm 120$ and $1210 \pm 300$ events for the three channels, respectively. 
Good agreement between data and MC simulation is observed for the overall normalization and the shapes of various kinematic distributions for events in this control region, 
within the statistical and systematic uncertainties, which are described below.

The $Z+$jets background is one of the dominant backgrounds in the $ee$
and $\mu\mu$ channels, and it is estimated using the data-driven method
described below, while its contribution in the $e\mu$ channel is found to be
small and estimated using the MC simulation.
This background is mainly due to 
mismeasurements of lepton or jet transverse momenta that result in large \met\ in the event.
Its contribution is suppressed by the high dilepton invariant mass and \met\ requirements. 
A control region dominated by $Z+$jets production (denoted by ``$Z+$jets control region'') is defined by 
applying the same set of selection cuts as for the signal region, 
but reversing the dilepton invariant mass cut to $60 < m_{\ell \ell} < 106$ GeV. 
Since the shape of the $m_{\ell \ell}$ distribution in data and MC simulation is in agreement over the full range $60 < m_{\ell \ell} < 1000$ GeV, 
the ratio $R$ of $Z+$jets events in the signal region to those in the control region is estimated using the MC simulation. 
The number of data events observed in the $Z+$jets control region, 
after having subtracted the non-$Z+$jets events contribution using MC expectations, 
is scaled by $R$ to estimate the $Z+$jets background contribution in the signal region. 
The ratio $R$ from $Z+$jets events generated with {\sc alpgen}, 
is found to be $0.040^{+ 0.005}_{- 0.006}$ in the $ee$ channel and $0.046^{+0.019}_{-0.015}$ in the $\mu\mu$ channel.  
The non-$Z+$jets events contribution in the $Z+$jets control region is
12\% in the $ee$ channel and 16\% in the $\mu\mu$ channel.  

The $W+$jets process contributes to the final selected sample when one or more hadrons in a jet decay to, or are
misidentified as, a charged lepton. Since the probability for a jet to be identified as a lepton may not be well modelled in the MC simulation, a data-driven method is used to estimate this contribution. 
A data control sample is selected by requiring one lepton which passes all the quality criteria in the lepton selection described above 
and a second lepton-like object. The muon-like objects are those reconstructed as muons but failing the isolation requirement. 
The electron-like objects are those reconstructed as {\it loose} electrons~\cite{electron_performance} but failing  both the isolation and the {\it tight} quality requirements.
These lepton-like objects are most likely jets reconstructed as leptons. 
To obtain the expected number of $W+$jets events contaminating the signal region,
the number of events in this $W+$jets dominated control sample is then scaled by a pass-to-fail ratio $f$,
defined as the number of lepton-like objects passing the full lepton selection requirements 
divided by the number that fail.
The non-$W+$jets events in the control region are subtracted using MC expectations. The factor $f$ is measured from data for electrons and muons separately, using two control samples dominated by di-jet events. The di-jet samples are selected by tagging events with one jet and one back-to-back lepton-like object without any isolation requirement (and no {\it tight} requirement for the electrons) after suppressing the lepton contribution from $W/Z$ bosons. The ratio $f$ is measured as a function of the jet \pt, and its value is found to be between 0.3 and 1.0 for electrons, and between 0.02 and 0.15 for muons. 

The background contribution from QCD di-jet events in the signal region is estimated in a similar way to the $W+$jets contribution, 
but in this case the control sample is selected by requiring two lepton-like jets, 
and the ratio $f$ is applied to both of them. This background contribution is found to be negligible. 

The simulation of the RS \G\ signal is based on the LO matrix element implemented in \pythia~\cite{pythia} 6.421 event generator, 
with the modified LO~\cite{Sherstnev:2008dm} parton distribution function (PDF) set MRST2007LO$^*$~\cite{Thorne:2009ky}. 
The coupling $\kappa/\bar{M}_{\rm pl}=0.1$ is assumed. 
A separate MC sample is generated for each of seven graviton masses $m_{\G} =$ 200, 350, 500, 750, 1000, 1250 and 1500 GeV. 
The production cross section times branching ratio 
$\sigma(pp\rightarrow \G) \times BR(\G \rightarrow WW \rightarrow \ell \nu \ell' \nu')$ ($\ell, \ell' = e, \mu$ or $\tau$) 
decreases from 108 pb to 1.8 fb when the simulated $m_{\G}$ increases from 200 GeV to 1500 GeV.
The \KKG\ signal is simulated at LO using {\sc calchep}~\cite{calchep}
v3.2 using the CTEQ6L1 PDF set~\cite{cteq6}, interfaced to {\sc
  pythia} for parton showering and hadronization. 
In order to compensate for the smaller production cross section with
respect to the original RS model, a larger coupling
$\kappa/\bar{M}_{\rm pl}=1.0$ is assumed when generating these
samples.  
Thirteen signal samples with \KKG\ masses between 300 GeV and 1500 GeV in 100 GeV mass steps are generated, 
with the predicted $\sigma(pp\rightarrow \KKG) \times BR(\KKG \rightarrow WW \rightarrow \ell \nu \ell' \nu')$ ($\ell, \ell' = e, \mu$ or $\tau$) decreasing from 8.6 pb to 0.22 fb. 
The ATLAS fast simulation~\cite{ATLASsim} is used to simulate the detector response for both \G\ and \KKG\ samples. 
Events with $W$ bosons decaying to $\tau$ leptons are also considered as part of the signal if electrons or muons are present in the final state. 
The overall acceptance times trigger, reconstruction and selection efficiencies ($A \times \epsilon$), 
defined as the number of signal events passing the full event selection divided by the number of generated events, 
increases from 3.0\% at $m_{\G}=200$ GeV to 40.9\% at $m_{\G}=1500$ GeV for \G. 
The corresponding $A \times \epsilon$ for $\KKG$ increases from 16.8\% at $m_{\KKG}=300$ GeV to 50.8\% at $m_{\KKG}=1500$ GeV. 
The difference in $A \times \epsilon$ between the two models is due to different production mechanisms and the treatment of the $W$ boson polarization in its decay, which is properly taken into account by {\sc calchep} but not by {\sc pythia}. 
{\sc pythia} is chosen to simulate the RS \G\ samples, even though it does not properly account for the $W$ boson polarization, in order to allow direct comparison  with previous search results, which used the same {\sc pythia} implementation to simulate this process.

Table~\ref{tab:expectedbkg-wsys} shows the number of events selected in data and
the estimated background contributions with combined statistical and systematic uncertainties. 
The expected numbers of events for an RS \G\ with a mass of 750 GeV and 1000 GeV, and for a \KKG\   
with a mass of 600 GeV and 1000 GeV are also reported.
A total of 1384 $\ell \ell' +\met$ candidates are observed in
data, while the expected number of events from SM processes is $1280 \pm 13({\rm stat}) \pm 200({\rm syst})$. 

Several sources of systematic uncertainty on the signal and background estimates are considered. 
The first is related to the correction scale factors applied to MC samples
in order to account for the difference in the performance of object reconstruction, 
identification, isolation and trigger efficiency between data and MC simulation. 
The uncertainty on the single-lepton trigger efficiency scale factor is 1\%, 
while the electron and muon reconstruction and identification efficiency scale factor uncertainties 
are less than 1.0\% and 0.4\% respectively, evaluated with tag-and-probe methods 
using $Z\rightarrow \ell\ell$, $W \rightarrow \ell\nu$  and $J/\psi \rightarrow \ell\ell$ events. 
A slight degradation of the muon reconstruction efficiency is observed at high \pt\ in simulated MC samples. 
An uncertainty of the order 1\% for muons with $\pt\ > 1$ \TeV, corresponding to the magnitude of this effect, is included.
The lepton isolation efficiency scale factor is determined with an uncertainty of 1\% and 0.3\% for electrons and muons, respectively.
The MC simulation is also corrected to reproduce the lepton energy scale and resolution, 
with residual uncertainties $< 1\%$ and $< 0.1\%$ on the energy scale, 
and $<0.6\%$ and $<5\%$ on the resolution, for electrons and muons, respectively. 
Uncertainties on the jet energy scale and resolution are found to be typically 3\% at high \met, relevant for this analysis, varying between 2-9\%~\cite{jetPerf}.
The uncertainties on the lepton and jet energy scale and resolution are propagated to the \met, which also receives contributions from energy deposits due to additional $pp$ collisions in the same or neighbouring bunch crossings, and from energy
deposits not associated to any reconstructed object. The total systematic uncertainty on the \met\ energy scale is 3.5\%~\cite{metPerf}.
The uncertainties on the $b$-tagging efficiency for heavy-quark jets and
mis-tag rate for light- and $c$-quark jets are measured in data, and
are 6$-$15\% and up to 21\%, respectively~\cite{btag}. 
The effect of all these sources of detector uncertainty on the shape of the distribution used to set the final cross-section limit is taken into account. 

The uncertainty on the normalization of the backgrounds estimated using MC simulation 
includes the integrated luminosity uncertainty of 3.9\%~\cite{Aad:2011dr,ATLAS-CONF-2011-116}, 
and the theoretical uncertainty on the inclusive cross sections of SM processes, 
namely 10\% for $\ttbar$~\cite{ttbar-xsec}, 9\% for single-top~\cite{st-xsec-1,st-xsec-2}, 5\% for $W/Z+$jets, 5\% for $WW$, 
7\% for $WZ$ and 5\% for $ZZ$~\cite{Campbell:2011bn}, which arises from the choice of PDFs, 
from factorization and renormalization scale dependence, and from strong coupling constant ($\alpha_{\rm s}$) variations.

The uncertainty on the estimate of the $W+$jets background includes
the uncertainty on the non-$W+$jets events subtraction in the control
region, and the uncertainty on the ratio $f$.  The uncertainty on the
non-$W+$jets background events is 10\%.  The uncertainty on
$f$ varies between 10\% and 30\%
depending on lepton \pt, and mainly comes from differences in the
kinematics and flavour composition of the di-jet events used to
determine the ratio $f$ with respect to the $W+$jets events to which
$f$ is applied. 

The uncertainty on the data-driven normalization of the $Z+$jets
background in the $ee$ and $\mu\mu$ due to the non-$Z+$jets events
subtraction in the control region is negligible, while the main
contribution comes from the uncertainty on the factor $R$. This is
evaluated accounting for possible uncertainties on the dilepton mass shape due to
initial and final state radiation modeling, and on the \met\ shape due
to parton shower and hadronization modeling, both determined using {\sc
  pythia} and {\sc alpgen} $Z+$jets simulations. The effect of
lepton scale and resolution, and \met\ resolution are also taken into account. 

Further systematic uncertainties on the \ttbar\ background are estimated,
including the difference between event generators, parton shower models and initial- and final-state radiation models. 
The dominant contribution (up to 40\%) is due to the parton shower model, arising from the $b$-jet requirement. 
The systematic uncertainties on the modelling of the kinematics of the SM $WW$ process have been evaluated by comparing different MC generators; 
the local differences in the distributions are found to be smaller than 10\%.

The effect on the signal acceptance due to the choice of the PDF set used to simulate the signal samples is also considered. It is estimated to be 1\% by comparing predictions of the nominal PDF set MRST2007 LO$^*$ with those of two NNPDF LO$^*$ 2.1~\cite{Ball:2010de} sets with values of $\alpha_{\rm s}=0.119,0.130$, and that of the CT09MCS~\cite{CT09MC2} PDF set, 
using the standard LHAPDF framework~\cite{Whalley:2005nh}. 

\begin{table*}
\begin{center}
\begin{tabular}{l|rrr|rrr|rrr} 
  \hline \hline

      Process               & \multicolumn{3}{c|}{$ee$}      &  \multicolumn{3}{c|}{$\mu\mu$}       &  \multicolumn{3}{c}{$e\mu$}    \\ \hline
      \WW\                  & $64.6$ & $\pm$ & $6.1$          & $82.3$ & $\pm$ & $6.8$                & $433$  & $\pm$ & $30$  \\
      $WZ$                  & $7.3$  & $\pm$ & $0.9$          & $7.7$  & $\pm$ & $0.9$                & $28.9$ & $\pm$ & $2.7$   \\
      $ZZ$                  & $2.7$  & $\pm$ & $0.4$          & $3.2$  & $\pm$ & $0.4$                & $1.5$  & $\pm$ & $0.3$   \\
      $W\gamma$             & $1.6$  & $\pm$ & $1.0$          & \multicolumn{3}{c|}{negl.}            & $7.6$  & $\pm$ & $2.4$    \\
      Single top            & $12.8$ & $\pm$ & $2.4$          & $16.7$ & $\pm$ & $2.7$                & $63$   & $\pm$ & $12$   \\
      $t\bar{t}$            & $59$   & $\pm$ & $31$           & $76$   & $\pm$ & $38$                 & $230$  & $\pm$ & $120$  \\
      $W+$jets              & $7.5$  & $\pm$ & $3.0$          & $4.7$  & $\pm$ & $1.9$                & $35.1$ & $\pm$ & $7.5$ \\
      $Z+$jets              & $55$   & $\pm$ & $10$           & $62$   & $\pm$ & $25$                 & $22.2$ & $\pm$ & $3.3$   \\  
      \hline
      Sum of all backgrounds       & $211$ & $\pm$ & $33$    & $253$ & $\pm$ & $46$                   & $820$ & $\pm$ & $120 $ \\
      Data                         & \multicolumn{3}{c|}{$258$}   & \multicolumn{3}{c|}{$249$}   & \multicolumn{3}{c}{$877$}   \\  \hline \hline
      RS \G\ ($m=750$ GeV)           & $28.9$ & $\pm$ & $1.7$   & $29.3$ & $\pm$ & $1.7$   & $73.0$ & $\pm$ & $3.9$ \\
      RS \G\ ($m=1000$ GeV)          & $6.4$ & $\pm$ & $0.4$   & $6.4$ & $\pm$ & $0.4$     & $15.3$ & $\pm$ & $0.8$ \\
      Bulk RS \G\ ($m=600$ GeV)       & $26.3$ & $\pm$ & $1.5$  & $25.7$ & $\pm$ & $1.5$    & $73.6$ & $\pm$ & $3.9$  \\
      Bulk RS \G\ ($m=1000$ GeV)      & $1.4$ & $\pm$ & $0.1$   & $1.2$ & $\pm$ & $0.1$     & $3.2$ & $\pm$ & $0.1$  \\ \hline \hline
\end{tabular}
\end{center} 
\caption{Estimated background yields, observed number of data events, and predicted signal yield for different graviton mass points for the three analysed channels. The quoted uncertainties are the combined statistical and systematic uncertainties.}
\label{tab:expectedbkg-wsys}
\end{table*}

No significant excess in the overall number of selected $WW$ events is observed in data. The transverse mass of the 
$WW$ candidates, defined as $m_{\rm T}^{WW} = \sqrt{ (\sum\limits_{i=1}^2 \pt^{\ell i} + \met)^2 -  (\sum\limits_{i=1}^2 p_x^{\ell i} +  E^{\rm miss}_x)^2 -  (\sum\limits_{i=1}^2 p_y^{\ell i} +  E^{\rm miss}_y)^2 }$, 
 is examined for any resonant structure, where $p_{{\rm T}(x, y)}^{\ell i}$ is the \pt\ ($p_x, p_y$) of the $i$-th lepton, and $ E^{\rm miss}_{x(y)}$ is the $x (y)$ component of the \met.
The \mtww\ distribution of the $WW$ system for the three analysed channels is presented in Fig.~\ref{fig:mtww}, for data and background expectations together with the expected signal contributions from RS graviton and bulk RS graviton models.  
Due to the small numbers of MC events, a convolution of a Gaussian with an exponential function is used to fit the \mtww\ distribution of each SM background. The functional form is then used to predict the background contribution in the region $\mtww\ > 300$ GeV, and the uncertainty on the fit parameters is treated as an additional systematic uncertainty on the final \mtww\ shape.

The \mtww\ distribution is used to build a log-likelihood ratio (LLR) test statistic~\cite{Junk:1999kv}
to assess the compatibility of the data with the presence of a signal in addition to the background in a modified frequentist approach~\cite{Cowan:2010js}. 
Confidence levels (CL) for the signal plus background
hypothesis, $\rm{CL}_{\rm s+b}$, and background-only hypothesis, $\rm{CL}_{\rm b}$, are computed by
integrating the LLR distributions obtained from simulated pseudo-experiments using Poisson statistics, and their ratio $\rm{CL}_{\rm s}$ is used to set the limits. 
Systematic uncertainties on the expected numbers of signal and background events are treated as nuisance parameters.
 The three analysed channels are treated separately and then combined
by summing up the LLR values over all bins. All correlations are
maintained among channels and between signal and background. 
Due to the large residual $Z+$jets background contamination in the $\mu\mu$ channel, caused by the worse muon resolution at high \pt, this channel has a poorer sensitivity than the other two channels.

No excess is observed in data and the $p$-value of the background-only hypothesis, defined as the probability for the background to produce an excess of equal or larger size than the observed one,  
is found to be greater than 0.08 in all \mtww\ regions. 
Upper limits are therefore derived on the production cross section times branching ratio ($\sigma \times BR$) for RS gravitons and bulk RS gravitons decaying to $WW$. 
The observed (expected) 95\% CL upper limits on $\sigma(pp \rightarrow \G /\KKG) \times BR(\G/\KKG \rightarrow WW)$ 
as a function of $m_{\G}$ and $m_{\KKG}$ are shown in Fig.~\ref{fig:1Dlimits} and reported in Tables~\ref{tab:rsg_xsection_limits} and~\ref{tab:kkg_xsection_limits}, 
corresponding to an observed (expected) 95\% CL lower limit of 1.23 (1.13) \TeV\ and 0.84 (0.74) \TeV\ on the masses of the \G\ and \KKG, respectively. 
Tables~\ref{tab:rsg_xsection_limits} and~\ref{tab:kkg_xsection_limits} also report the $A \times \epsilon$  values for each signal sample.

\begin{table}
\begin{center}
\begin{tabular}{cccc}
\hline \hline
 $m_{\G}$ [\GeV] & $A \times \epsilon$ [\%] & Expected [pb] & Observed [pb] \\ \hline
200 & 3.0 $\pm$ 0.1 & 17.6 & 20.3 \\
350 & 16.8 $\pm$ 0.5 & 4.68 & 5.51 \\
500 & 24.4 $\pm$ 0.7 & 1.30 & 1.46 \\
750 & 30.7 $\pm$ 0.9 & 0.315 & 0.264 \\
1000 & 36.3 $\pm$ 1.0 & 0.130 & 0.084 \\
1250 & 39.0 $\pm$ 1.1 & 0.085 & 0.062 \\
1500 & 40.9 $\pm$ 1.1 & 0.079 & 0.061 \\
\hline \hline
\end{tabular}
\end{center}
\caption{Expected and observed 95\% CL upper limits on the cross section times branching ratio $\sigma(pp
  \rightarrow \G\ ) \times Br(\G \rightarrow WW)$ as a function of the
  RS graviton mass. For each mass point, $A \times \epsilon$ is
  also reported with the combined statistical and systematic uncertainty.}
\label{tab:rsg_xsection_limits}
\end{table}

\begin{table}
\begin{center}
\begin{tabular}{cccc}
\hline \hline
 $m_{\KKG}$ [\GeV] &$A \times \epsilon$ [\%] & Expected [pb] & Observed [pb] \vspace{0.5mm} \\ \hline
300 & 16.8 $\pm$ 0.5 & 4.73 & 5.48 \\
400 & 26.5 $\pm$ 0.8 & 1.81 & 2.13 \\
500 & 33.6 $\pm$ 1.0 & 0.814 & 0.910 \\
600 & 39.0 $\pm$ 1.1 & 0.398 & 0.405 \\
700 & 42.3 $\pm$ 1.2 & 0.212 & 0.189 \\
800 & 44.2 $\pm$ 1.2 & 0.134 & 0.102 \\
900 & 46.1 $\pm$ 1.3 & 0.083 & 0.056 \\
1000 & 47.3 $\pm$ 1.3 & 0.060 & 0.040 \\
1100 & 48.9 $\pm$ 1.4 & 0.044 & 0.029 \\
1200 & 49.2 $\pm$ 1.4 & 0.037 & 0.025 \\
1300 & 50.1 $\pm$ 1.4 & 0.030 & 0.022 \\
1400 & 50.4 $\pm$ 1.4 & 0.028 & 0.019 \\
1500 & 50.8 $\pm$ 1.4 & 0.027 & 0.020 \\
\hline \hline
\end{tabular}
\end{center}
\caption{Expected and observed 95\% CL upper limits on the cross section times branching ratio $\sigma(pp
  \rightarrow \KKG\ ) \times Br(\KKG \rightarrow WW)$ as a function of
  the bulk RS graviton mass. For each mass point, $A \times \epsilon$ is
  also reported with the combined statistical and systematic uncertainty.}
\label{tab:kkg_xsection_limits}
\end{table}

\begin{figure*}
  \centering
  \subfigure[]{\includegraphics[width=0.45\textwidth]{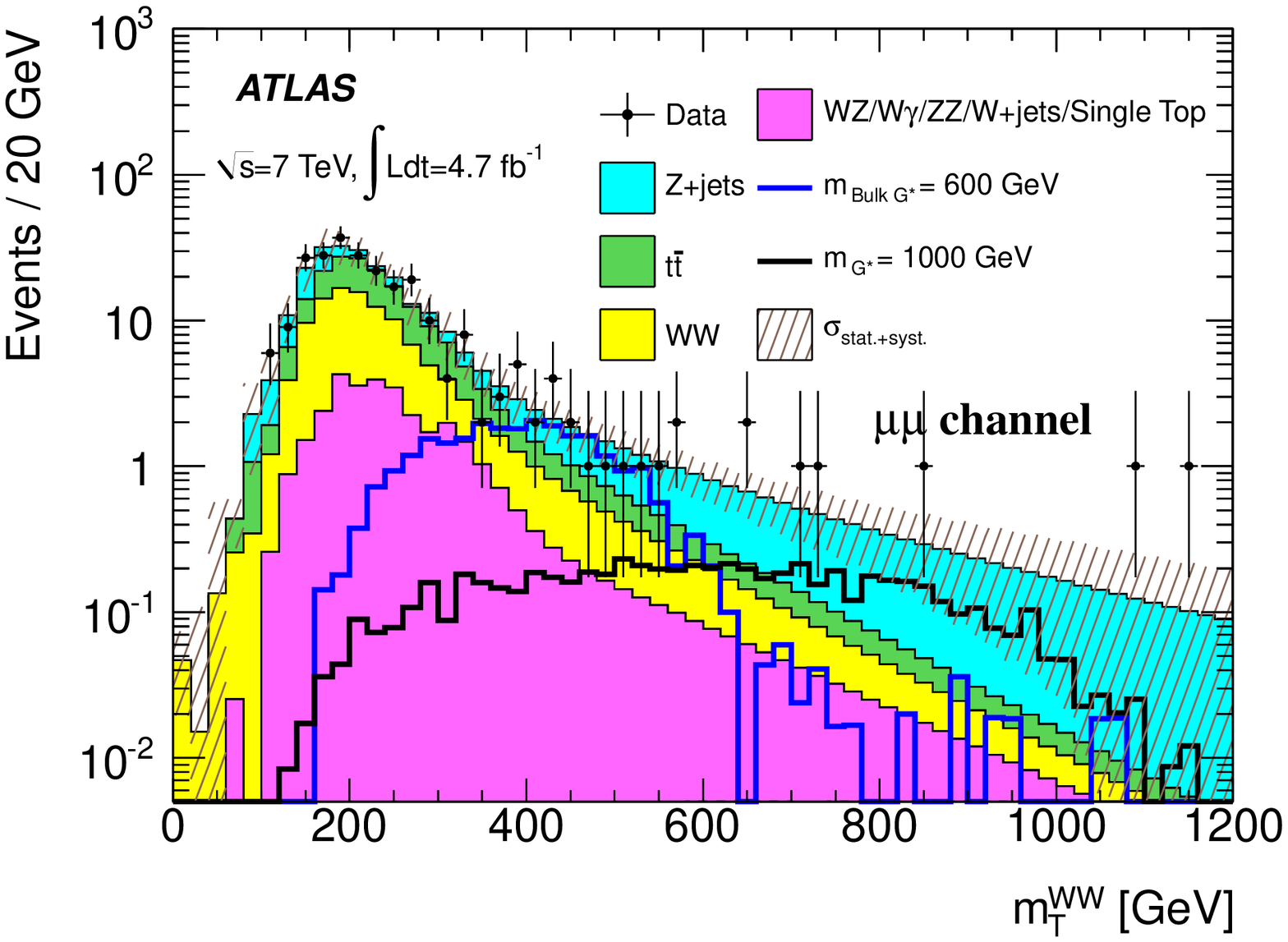}\label{f-mtww-mm}}  
  \subfigure[]{\includegraphics[width=0.45\textwidth]{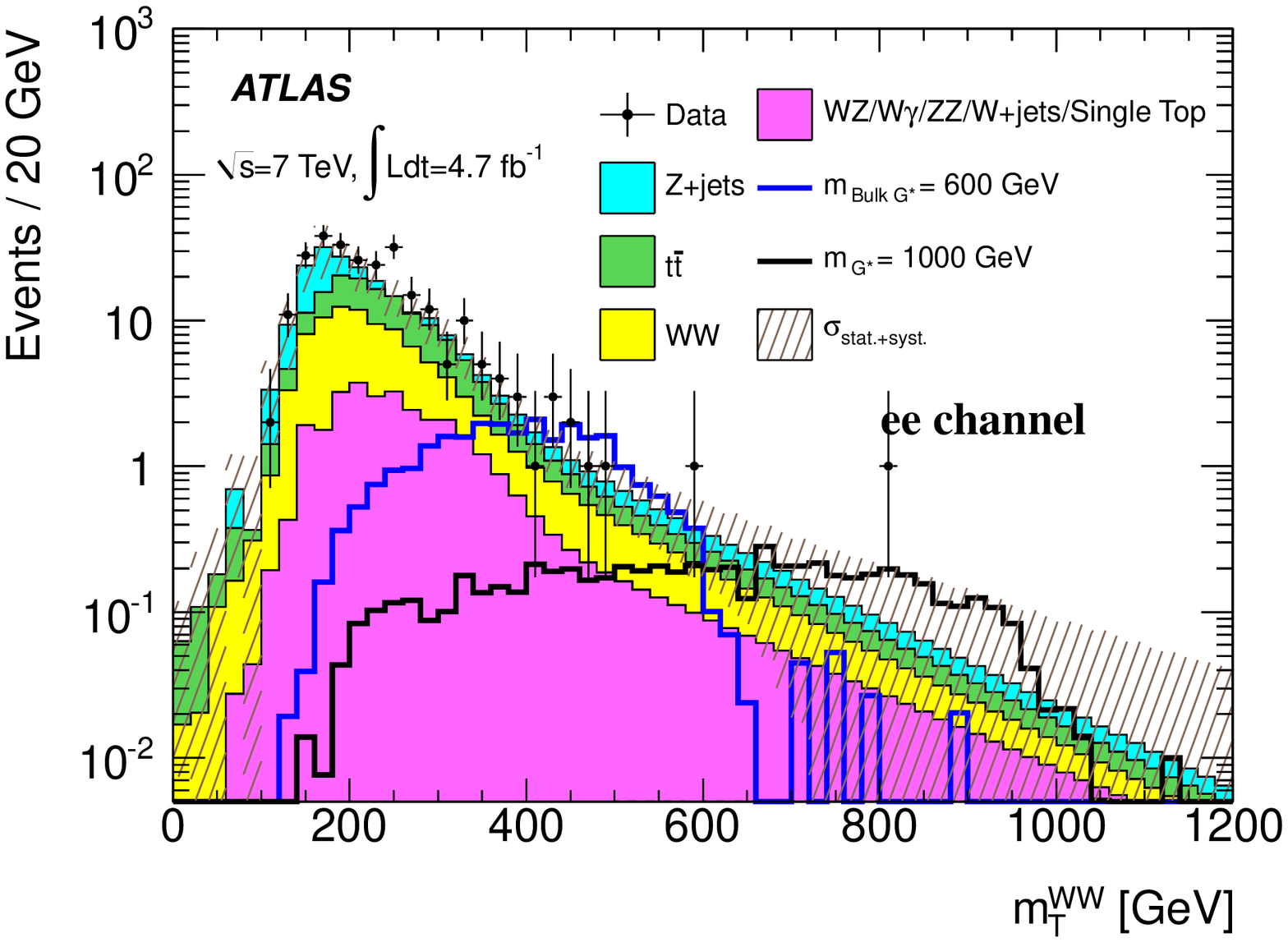}\label{f-mtww-ee}}  \\
  \subfigure[]{\includegraphics[width=0.45\textwidth]{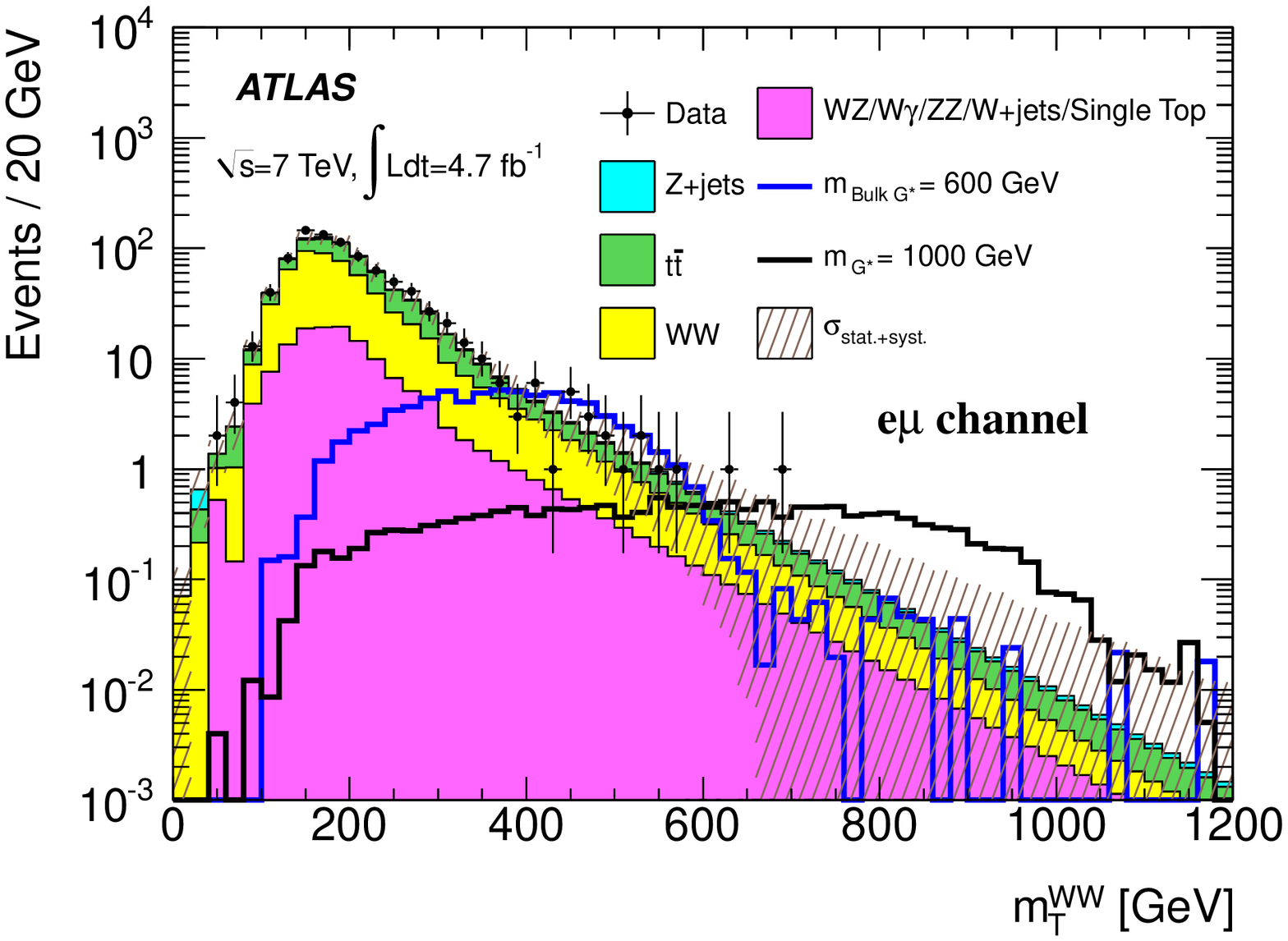}\label{f-mtww-em}}  
  \caption{Observed and predicted \mtww\ distribution after event selection in the (a) $\mu\mu$, (b) $ee$ and (c) $e\mu$ channels. 
           For $\mtww>300$ GeV, the predicted backgrounds are obtained from fits to the MC samples. 
           Predictions for an RS graviton with a mass of 1000 GeV and a bulk RS graviton with a mass of 600 GeV are also shown. 
           The shaded area represents the total statistical and systematic uncertainty on the background prediction.     
  \label{fig:mtww}}
\end{figure*}

\begin{figure*}
\centering
  \subfigure[]{\includegraphics[width=0.45\textwidth]{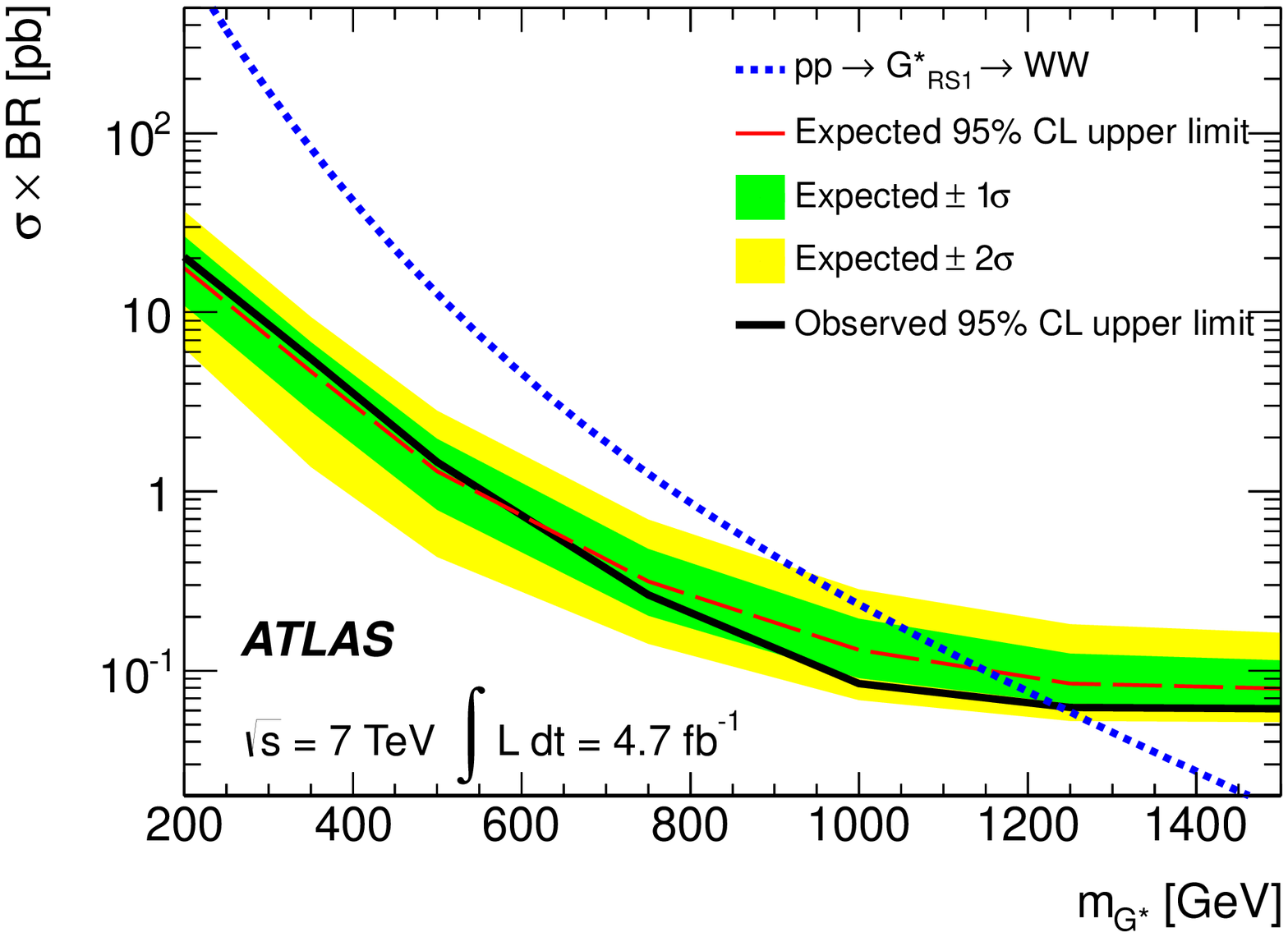}\label{f-wprime-limit}} 
  \subfigure[]{\includegraphics[width=0.45\textwidth]{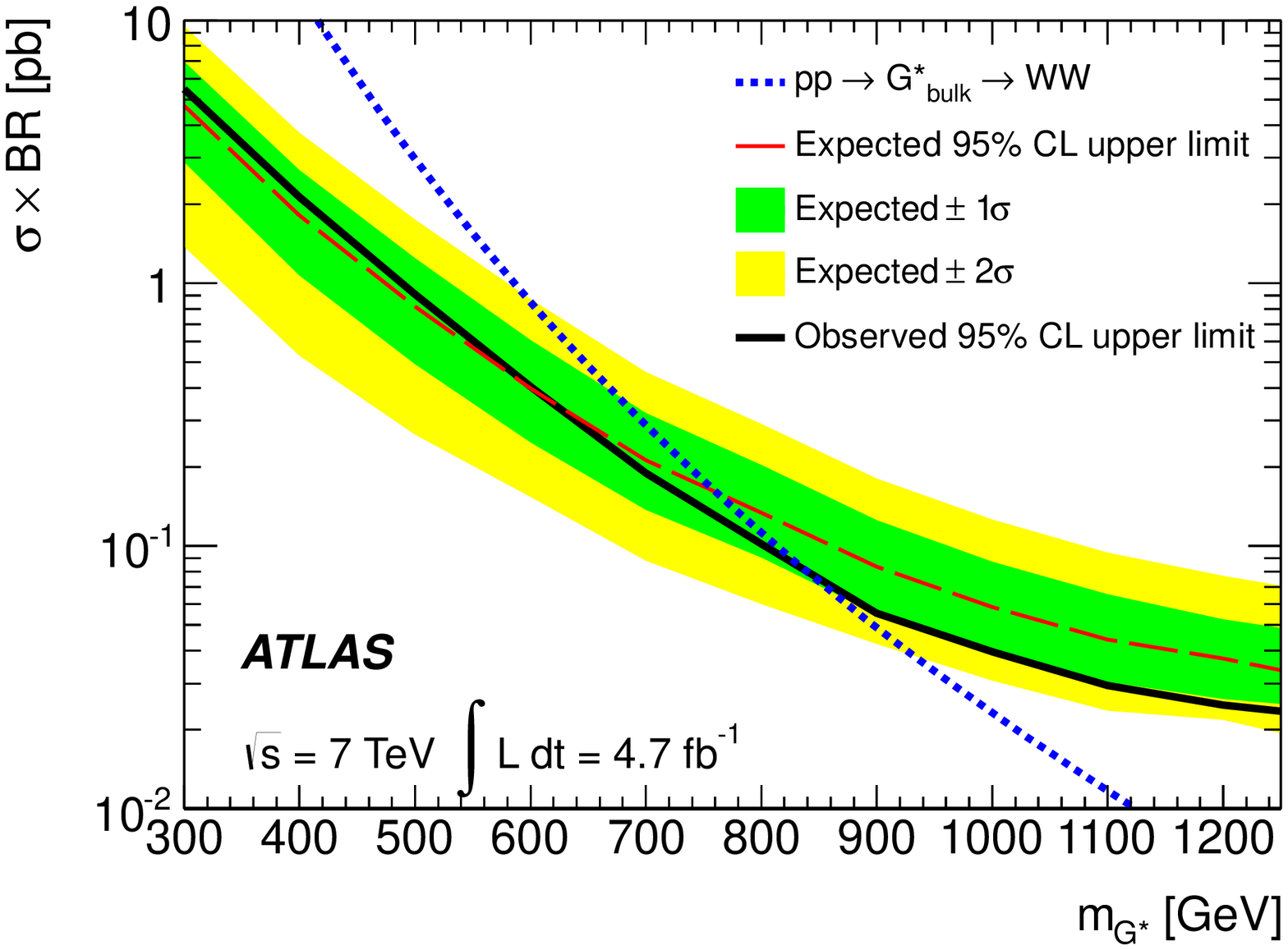}\label{f-technirho-limit}}
 \caption{
    The observed and expected 95\% CL upper limits on $\sigma \times BR$ 
    for (a) the RS graviton $\sigma(pp \rightarrow \G) \times BR(\G \rightarrow WW)$ and 
    (b) the bulk RS graviton $\sigma(pp \rightarrow \KKG) \times BR(\KKG \rightarrow WW)$, 
    with the theoretical predictions at LO (dotted line). The inner and outer bands represent respectively the $1\sigma$ and $2\sigma$ uncertainty on the expected limit.
 \label{fig:1Dlimits}}
\end{figure*}

In conclusion, a generic search for resonant production of a pair of $W$ bosons in two opposite sign leptons and 
large \met\ final state has been performed using 4.7 fb$^{-1}$ of data 
collected with the ATLAS detector in $pp$ collisions at $\sqrt{s}=7$ \TeV~at the LHC.
No significant excess of events is observed and upper limits on the production cross section times branching ratio 
are set for two benchmark models: RS $G^*$ and bulk RS $G^*$.
The observed (expected) 95\% CL lower limit on the masses of the two
particles is found to be 1.23 (1.13) \TeV\ for \G\ and 0.84 (0.74)
\TeV\ for \KKG, assuming the coupling $\kappa/ \bar{M}_{\rm pl}=0.1$
and $\kappa/ \bar{M}_{\rm pl}=1.0$, respectively.


We thank CERN for the very successful operation of the LHC, as well as the
support staff from our institutions without whom ATLAS could not be
operated efficiently.

We acknowledge the support of ANPCyT, Argentina; YerPhI, Armenia; ARC,
Australia; BMWF, Austria; ANAS, Azerbaijan; SSTC, Belarus; CNPq and FAPESP,
Brazil; NSERC, NRC and CFI, Canada; CERN; CONICYT, Chile; CAS, MOST and NSFC,
China; COLCIENCIAS, Colombia; MSMT CR, MPO CR and VSC CR, Czech Republic;
DNRF, DNSRC and Lundbeck Foundation, Denmark; EPLANET and ERC, European Union;
IN2P3-CNRS, CEA-DSM/IRFU, France; GNSF, Georgia; BMBF, DFG, HGF, MPG and AvH
Foundation, Germany; GSRT, Greece; ISF, MINERVA, GIF, DIP and Benoziyo Center,
Israel; INFN, Italy; MEXT and JSPS, Japan; CNRST, Morocco; FOM and NWO,
Netherlands; RCN, Norway; MNiSW, Poland; GRICES and FCT, Portugal; MERYS
(MECTS), Romania; MES of Russia and ROSATOM, Russian Federation; JINR; MSTD,
Serbia; MSSR, Slovakia; ARRS and MVZT, Slovenia; DST/NRF, South Africa;
MICINN, Spain; SRC and Wallenberg Foundation, Sweden; SER, SNSF and Cantons of
Bern and Geneva, Switzerland; NSC, Taiwan; TAEK, Turkey; STFC, the Royal
Society and Leverhulme Trust, United Kingdom; DOE and NSF, United States of
America.

The crucial computing support from all WLCG partners is acknowledged
gratefully, in particular from CERN and the ATLAS Tier-1 facilities at
TRIUMF (Canada), NDGF (Denmark, Norway, Sweden), CC-IN2P3 (France),
KIT/GridKA (Germany), INFN-CNAF (Italy), NL-T1 (Netherlands), PIC (Spain),
ASGC (Taiwan), RAL (UK) and BNL (USA) and in the Tier-2 facilities
worldwide.

\bibliographystyle{apsrev4-1}
\bibliography{WWlvlv}

\begin{thebibliography}{46}%
\makeatletter
\providecommand \@ifxundefined [1]{%
 \@ifx{#1\undefined}
}%
\providecommand \@ifnum [1]{%
 \ifnum #1\expandafter \@firstoftwo
 \else \expandafter \@secondoftwo
 \fi
}%
\providecommand \@ifx [1]{%
 \ifx #1\expandafter \@firstoftwo
 \else \expandafter \@secondoftwo
 \fi
}%
\providecommand \natexlab [1]{#1}%
\providecommand \enquote  [1]{``#1''}%
\providecommand \bibnamefont  [1]{#1}%
\providecommand \bibfnamefont [1]{#1}%
\providecommand \citenamefont [1]{#1}%
\providecommand \href@noop [0]{\@secondoftwo}%
\providecommand \href [0]{\begingroup \@sanitize@url \@href}%
\providecommand \@href[1]{\@@startlink{#1}\@@href}%
\providecommand \@@href[1]{\endgroup#1\@@endlink}%
\providecommand \@sanitize@url [0]{\catcode `\\12\catcode `\$12\catcode
  `\&12\catcode `\#12\catcode `\^12\catcode `\_12\catcode `\%12\relax}%
\providecommand \@@startlink[1]{}%
\providecommand \@@endlink[0]{}%
\providecommand \url  [0]{\begingroup\@sanitize@url \@url }%
\providecommand \@url [1]{\endgroup\@href {#1}{\urlprefix }}%
\providecommand \urlprefix  [0]{URL }%
\providecommand \Eprint [0]{\href }%
\providecommand \doibase [0]{http://dx.doi.org/}%
\providecommand \selectlanguage [0]{\@gobble}%
\providecommand \bibinfo  [0]{\@secondoftwo}%
\providecommand \bibfield  [0]{\@secondoftwo}%
\providecommand \translation [1]{[#1]}%
\providecommand \BibitemOpen [0]{}%
\providecommand \bibitemStop [0]{}%
\providecommand \bibitemNoStop [0]{.\EOS\space}%
\providecommand \EOS [0]{\spacefactor3000\relax}%
\providecommand \BibitemShut  [1]{\csname bibitem#1\endcsname}%
\let\auto@bib@innerbib\@empty
\bibitem [{\citenamefont {Altarelli}\ \emph {et~al.}(1989)\citenamefont
  {Altarelli}, \citenamefont {Mele},\ and\ \citenamefont {Ruiz-Altaba}}]{egm}%
  \BibitemOpen
  \bibfield  {author} {\bibinfo {author} {\bibfnamefont {G.}~\bibnamefont
  {Altarelli}}, \bibinfo {author} {\bibfnamefont {B.}~\bibnamefont {Mele}}, \
  and\ \bibinfo {author} {\bibfnamefont {M.}~\bibnamefont {Ruiz-Altaba}},\
  }\href@noop {} {\bibfield  {journal} {\bibinfo  {journal} {Z. Phys. C -
  Particles and Fields}\ }\textbf {\bibinfo {volume} {45}},\ \bibinfo {pages}
  {109} (\bibinfo {year} {1989})}\BibitemShut {NoStop}%
\bibitem [{\citenamefont {Randall}\ and\ \citenamefont
  {Sundrum}(1999)}]{extraDimension1}%
  \BibitemOpen
  \bibfield  {author} {\bibinfo {author} {\bibfnamefont {L.}~\bibnamefont
  {Randall}}\ and\ \bibinfo {author} {\bibfnamefont {R.}~\bibnamefont
  {Sundrum}},\ }\href@noop {} {\bibfield  {journal} {\bibinfo  {journal} {Phys.
  Rev. Lett.}\ }\textbf {\bibinfo {volume} {83}},\ \bibinfo {pages} {3370}
  (\bibinfo {year} {1999})},\ \Eprint
  {http://arxiv.org/abs/hep-ph/9905221}{arXiv:hep-ph/9905221
  [hep-ph]}\BibitemShut {NoStop}%
\bibitem [{\citenamefont {Davoudiasl}\ \emph {et~al.}(2000)\citenamefont
  {Davoudiasl}, \citenamefont {Hewett},\ and\ \citenamefont
  {Rizzo}}]{extraDimension2}%
  \BibitemOpen
  \bibfield  {author} {\bibinfo {author} {\bibfnamefont {H.}~\bibnamefont
  {Davoudiasl}}, \bibinfo {author} {\bibfnamefont {J.~L.}\ \bibnamefont
  {Hewett}}, \ and\ \bibinfo {author} {\bibfnamefont {T.~G.}\ \bibnamefont
  {Rizzo}},\ }\href@noop {} {\bibfield  {journal} {\bibinfo  {journal} {Phys.
  Lett. B}\ }\textbf {\bibinfo {volume} {473}},\ \bibinfo {pages} {43}
  (\bibinfo {year} {2000})},\ \Eprint
  {http://arxiv.org/abs/hep-ph/9911262}{arXiv:hep-ph/9911262
  [hep-ph]}\BibitemShut {NoStop}%
\bibitem [{\citenamefont {Arkani-Hamed}\ \emph {et~al.}(1998)\citenamefont
  {Arkani-Hamed}, \citenamefont {Dimopoulos},\ and\ \citenamefont
  {Dvali}}]{extraDimension3}%
  \BibitemOpen
  \bibfield  {author} {\bibinfo {author} {\bibfnamefont {N.}~\bibnamefont
  {Arkani-Hamed}}, \bibinfo {author} {\bibfnamefont {S.}~\bibnamefont
  {Dimopoulos}}, \ and\ \bibinfo {author} {\bibfnamefont {G.}~\bibnamefont
  {Dvali}},\ }\href@noop {} {\bibfield  {journal} {\bibinfo  {journal} {Phys.
  Lett.}\ }\textbf {\bibinfo {volume} {B429}},\ \bibinfo {pages} {263}
  (\bibinfo {year} {1998})},\ \Eprint
  {http://arxiv.org/abs/hep-ph/9803315}{arXiv:hep-ph/9803315
  [hep-ph]}\BibitemShut {NoStop}%
\bibitem [{\citenamefont {Antoniadis}\ \emph {et~al.}(1998)\citenamefont
  {Antoniadis}, \citenamefont {Arkani-Hamed}, \citenamefont {Dimopoulos},\ and\
  \citenamefont {Dvali}}]{extraDimension4}%
  \BibitemOpen
  \bibfield  {author} {\bibinfo {author} {\bibfnamefont {I.}~\bibnamefont
  {Antoniadis}}, \bibinfo {author} {\bibfnamefont {N.}~\bibnamefont
  {Arkani-Hamed}}, \bibinfo {author} {\bibfnamefont {S.}~\bibnamefont
  {Dimopoulos}}, \ and\ \bibinfo {author} {\bibfnamefont {G.}~\bibnamefont
  {Dvali}},\ }\href@noop {} {\bibfield  {journal} {\bibinfo  {journal} {Phys.
  Lett.}\ }\textbf {\bibinfo {volume} {B436}},\ \bibinfo {pages} {257}
  (\bibinfo {year} {1998})},\ \Eprint
  {http://arxiv.org/abs/hep-ph/9804398}{arXiv:hep-ph/9804398
  [hep-ph]}\BibitemShut {NoStop}%
\bibitem [{\citenamefont {Arkani-Hamed}\ \emph {et~al.}(1999)\citenamefont
  {Arkani-Hamed}, \citenamefont {Dimopoulos},\ and\ \citenamefont
  {Dvali}}]{extraDimension5}%
  \BibitemOpen
  \bibfield  {author} {\bibinfo {author} {\bibfnamefont {N.}~\bibnamefont
  {Arkani-Hamed}}, \bibinfo {author} {\bibfnamefont {S.}~\bibnamefont
  {Dimopoulos}}, \ and\ \bibinfo {author} {\bibfnamefont {G.}~\bibnamefont
  {Dvali}},\ }\href {\doibase 10.1103/PhysRevD.59.086004} {\bibfield  {journal}
  {\bibinfo  {journal} {Phys. Rev.}\ }\textbf {\bibinfo {volume} {D59}},\
  \bibinfo {pages} {086004} (\bibinfo {year} {1999})},\ \Eprint
  {http://arxiv.org/abs/hep-ph/9807344}{arXiv:hep-ph/9807344
  [hep-ph]}\BibitemShut {NoStop}%
\bibitem [{\citenamefont {Eichten}\ and\ \citenamefont {Lane}(2008)}]{TC1}%
  \BibitemOpen
  \bibfield  {author} {\bibinfo {author} {\bibfnamefont {E.}~\bibnamefont
  {Eichten}}\ and\ \bibinfo {author} {\bibfnamefont {K.}~\bibnamefont {Lane}},\
  }\href {\doibase 10.1016/j.physletb.2008.09.047} {\bibfield  {journal}
  {\bibinfo  {journal} {Phys. Lett.}\ }\textbf {\bibinfo {volume} {B669}},\
  \bibinfo {pages} {235} (\bibinfo {year} {2008})},\ \Eprint
  {http://arxiv.org/abs/0706.2339}{arXiv:0706.2339 [hep-ph]}\BibitemShut
  {NoStop}%
\bibitem [{\citenamefont {Catterall}\ \emph {et~al.}(2012)\citenamefont
  {Catterall}, \citenamefont {Del~Debbio}, \citenamefont {Giedt},\ and\
  \citenamefont {Keegan}}]{TC2}%
  \BibitemOpen
  \bibfield  {author} {\bibinfo {author} {\bibfnamefont {S.}~\bibnamefont
  {Catterall}}, \bibinfo {author} {\bibfnamefont {L.}~\bibnamefont
  {Del~Debbio}}, \bibinfo {author} {\bibfnamefont {J.}~\bibnamefont {Giedt}}, \
  and\ \bibinfo {author} {\bibfnamefont {L.}~\bibnamefont {Keegan}},\
  }\href@noop {} {\bibfield  {journal} {\bibinfo  {journal} {Phys. Rev.}\
  }\textbf {\bibinfo {volume} {D85}},\ \bibinfo {pages} {094501} (\bibinfo
  {year} {2012})},\ \Eprint {http://arxiv.org/abs/1108.3794}{arXiv:1108.3794
  [hep-ph]}\BibitemShut {NoStop}%
\bibitem [{\citenamefont {Andersen}\ \emph {et~al.}(2011)\citenamefont
  {Andersen} \emph {et~al.}}]{TC3}%
  \BibitemOpen
  \bibfield  {author} {\bibinfo {author} {\bibfnamefont {J.}~\bibnamefont
  {Andersen}} \emph {et~al.},\ }\href@noop {} {\bibfield  {journal} {\bibinfo
  {journal} {Eur. Phys. J.}\ }\textbf {\bibinfo {volume} {126}},\ \bibinfo
  {pages} {81} (\bibinfo {year} {2011})},\ \Eprint
  {http://arxiv.org/abs/1104.1255}{arXiv:1104.1255 [hep-ph]}\BibitemShut
  {NoStop}%
\bibitem [{\citenamefont {Agashe}\ \emph {et~al.}(2007)\citenamefont {Agashe},
  \citenamefont {Davoudiasl}, \citenamefont {Perez},\ and\ \citenamefont
  {Soni}}]{brs_model}%
  \BibitemOpen
  \bibfield  {author} {\bibinfo {author} {\bibfnamefont {K.}~\bibnamefont
  {Agashe}}, \bibinfo {author} {\bibfnamefont {H.}~\bibnamefont {Davoudiasl}},
  \bibinfo {author} {\bibfnamefont {G.}~\bibnamefont {Perez}}, \ and\ \bibinfo
  {author} {\bibfnamefont {A.}~\bibnamefont {Soni}},\ }\href@noop {} {\bibfield
   {journal} {\bibinfo  {journal} {Phys. Rev D}\ }\textbf {\bibinfo {volume}
  {76}},\ \bibinfo {pages} {036006} (\bibinfo {year} {2007})},\ \Eprint
  {http://arxiv.org/abs/0701186}{arXiv:0701186 [hep-ph]}\BibitemShut {NoStop}%
\bibitem [{\citenamefont {Abazov}\ \emph {et~al.}(2011)\citenamefont {Abazov}
  \emph {et~al.}}]{D0_WZ}%
  \BibitemOpen
  \bibfield  {author} {\bibinfo {author} {\bibfnamefont {V.~M.}\ \bibnamefont
  {Abazov}} \emph {et~al.} (\bibinfo {collaboration} {D0}),\ }\href@noop {}
  {\bibfield  {journal} {\bibinfo  {journal} {Phys. Rev. Lett.}\ }\textbf
  {\bibinfo {volume} {107}},\ \bibinfo {pages} {011801} (\bibinfo {year}
  {2011})},\ \Eprint {http://arxiv.org/abs/1011.6278}{arXiv:1011.6278
  [hep-ex]}\BibitemShut {NoStop}%
\bibitem [{\citenamefont {Aaltonen}\ \emph {et~al.}(2010)\citenamefont
  {Aaltonen} \emph {et~al.}}]{CDF_WW}%
  \BibitemOpen
  \bibfield  {author} {\bibinfo {author} {\bibfnamefont {T.}~\bibnamefont
  {Aaltonen}} \emph {et~al.} (\bibinfo {collaboration} {CDF}),\ }\href@noop {}
  {\bibfield  {journal} {\bibinfo  {journal} {Phys. Rev. Lett.}\ }\textbf
  {\bibinfo {volume} {104}},\ \bibinfo {pages} {241801} (\bibinfo {year}
  {2010})},\ \Eprint {http://arxiv.org/abs/1004.4946}{arXiv:1004.4946
  [hep-ex]}\BibitemShut {NoStop}%
\bibitem [{\citenamefont {{ATLAS Collaboration}}(2008)}]{atlas_detector}%
  \BibitemOpen
  \bibfield  {author} {\bibinfo {author} {\bibnamefont {{ATLAS
  Collaboration}}},\ }\href {\doibase 10.1088/1748-0221/3/08/S08003} {\bibfield
   {journal} {\bibinfo  {journal} {JINST}\ }\textbf {\bibinfo {volume} {3}},\
  \bibinfo {pages} {S08003} (\bibinfo {year} {2008})}\BibitemShut {NoStop}%
\bibitem [{geo()}]{geometry}%
  \BibitemOpen
  \href@noop {} {}\bibinfo {note} {ATLAS uses a right-handed coordinate system
  with its origin at the nominal interaction point~(IP) in the centre of the
  detector and the $z$-axis along the beam pipe. The $x$-axis points from the
  IP to the centre of the LHC ring, and the $y$ axis points upward. Cylindrical
  coordinates $(R,\phi)$ are used in the transverse plane, $\phi$ being the
  azimuthal angle around the beam pipe. The pseudorapidity is defined in terms
  of the polar angle $\theta$ as $\eta=-\ln\tan(\theta/2)$.}\BibitemShut
  {Stop}%
\bibitem [{\citenamefont {{ATLAS
  Collaboration}}(2011{\natexlab{a}})}]{Aad:2011dr}%
  \BibitemOpen
  \bibfield  {author} {\bibinfo {author} {\bibnamefont {{ATLAS
  Collaboration}}},\ }\href@noop {} {\bibfield  {journal} {\bibinfo  {journal}
  {Eur. Phys. J.}\ }\textbf {\bibinfo {volume} {C71}},\ \bibinfo {pages} {1630}
  (\bibinfo {year} {2011}{\natexlab{a}})},\ \Eprint
  {http://arxiv.org/abs/1101.2185}{arXiv:1101.2185 [hep-ex]}\BibitemShut
  {NoStop}%
\bibitem [{\citenamefont {{ATLAS
  Collaboration}}(2011{\natexlab{b}})}]{ATLAS-CONF-2011-116}%
  \BibitemOpen
  \bibfield  {author} {\bibinfo {author} {\bibnamefont {{ATLAS
  Collaboration}}},\ }\href@noop {} {\bibfield  {journal} {\bibinfo  {journal}
  {ATLAS-CONF-2011-116, available at https://cdsweb.cern.ch/record/1367408}\ }
  (\bibinfo {year} {2011}{\natexlab{b}})}\BibitemShut {NoStop}%
\bibitem [{\citenamefont {{ATLAS
  Collaboration}}(2012{\natexlab{a}})}]{electron_performance}%
  \BibitemOpen
  \bibfield  {author} {\bibinfo {author} {\bibnamefont {{ATLAS
  Collaboration}}},\ }\href@noop {} {\bibfield  {journal} {\bibinfo  {journal}
  {Eur. Phys. J.}\ }\textbf {\bibinfo {volume} {C72}},\ \bibinfo {pages} {1909}
  (\bibinfo {year} {2012}{\natexlab{a}})},\ \Eprint
  {http://arxiv.org/abs/1110.3174}{arXiv:1110.3174 [hep-ex]}\BibitemShut
  {NoStop}%
\bibitem [{\citenamefont {{ATLAS
  Collaboration}}(2011{\natexlab{c}})}]{jetPerf}%
  \BibitemOpen
  \bibfield  {author} {\bibinfo {author} {\bibnamefont {{ATLAS
  Collaboration}}},\ }\href@noop {} {\  (\bibinfo {year}
  {2011}{\natexlab{c}})},\ \Eprint
  {http://arxiv.org/abs/1112.6426}{arXiv:1112.6426 [hep-ex]}\BibitemShut
  {NoStop}%
\bibitem [{\citenamefont {Cacciari}\ \emph {et~al.}(2008)\citenamefont
  {Cacciari}, \citenamefont {Salam},\ and\ \citenamefont {Soyez}}]{antikt}%
  \BibitemOpen
  \bibfield  {author} {\bibinfo {author} {\bibfnamefont {M.}~\bibnamefont
  {Cacciari}}, \bibinfo {author} {\bibfnamefont {G.~P.}\ \bibnamefont {Salam}},
  \ and\ \bibinfo {author} {\bibfnamefont {G.}~\bibnamefont {Soyez}},\ }\href
  {\doibase 10.1088/1126-6708/2008/04/063} {\bibfield  {journal} {\bibinfo
  {journal} {JHEP}\ }\textbf {\bibinfo {volume} {0804}},\ \bibinfo {pages}
  {063} (\bibinfo {year} {2008})},\ \Eprint
  {http://arxiv.org/abs/0802.1189}{arXiv:0802.1189 [hep-ph]}\BibitemShut
  {NoStop}%
\bibitem [{\citenamefont {{ATLAS
  Collaboration}}(2011{\natexlab{d}})}]{jet_study}%
  \BibitemOpen
  \bibfield  {author} {\bibinfo {author} {\bibnamefont {{ATLAS
  Collaboration}}},\ }\href@noop {} {\bibfield  {journal} {\bibinfo  {journal}
  {ATLAS-CONF-2011-032, available at https://cdsweb.cern.ch/record/1333972}\ }
  (\bibinfo {year} {2011}{\natexlab{d}})}\BibitemShut {NoStop}%
\bibitem [{\citenamefont {{ATLAS Collaboration}}(2011{\natexlab{e}})}]{btag}%
  \BibitemOpen
  \bibfield  {author} {\bibinfo {author} {\bibnamefont {{ATLAS
  Collaboration}}},\ }\href@noop {} {\bibfield  {journal} {\bibinfo  {journal}
  {ATLAS-CONF-2011-102, available at http://cdsweb.cern.ch/record/1369219}\ }
  (\bibinfo {year} {2011}{\natexlab{e}})}\BibitemShut {NoStop}%
\bibitem [{\citenamefont {{ATLAS Collaboration}}(2012{\natexlab{b}})}]{met}%
  \BibitemOpen
  \bibfield  {author} {\bibinfo {author} {\bibnamefont {{ATLAS
  Collaboration}}},\ }\href@noop {} {\bibfield  {journal} {\bibinfo  {journal}
  {Eur. Phys. J.}\ }\textbf {\bibinfo {volume} {C72}},\ \bibinfo {pages} {1844}
  (\bibinfo {year} {2012}{\natexlab{b}})},\ \Eprint
  {http://arxiv.org/abs/1108.5602}{arXiv:1108.5602 [hep-ex]}\BibitemShut
  {NoStop}%
\bibitem [{\citenamefont {{ATLAS
  Collaboration}}(2010{\natexlab{a}})}]{atlas_simulation}%
  \BibitemOpen
  \bibfield  {author} {\bibinfo {author} {\bibnamefont {{ATLAS
  Collaboration}}},\ }\href@noop {} {\bibfield  {journal} {\bibinfo  {journal}
  {Eur. Phys. J.}\ }\textbf {\bibinfo {volume} {C70}},\ \bibinfo {pages} {823}
  (\bibinfo {year} {2010}{\natexlab{a}})},\ \Eprint
  {http://arxiv.org/abs/1005.4568}{arXiv:1005.4568}\BibitemShut {NoStop}%
\bibitem [{\citenamefont {Agostinelli}\ \emph {et~al.}(2003)\citenamefont
  {Agostinelli} \emph {et~al.}}]{geant}%
  \BibitemOpen
  \bibfield  {author} {\bibinfo {author} {\bibfnamefont {S.}~\bibnamefont
  {Agostinelli}} \emph {et~al.} (\bibinfo {collaboration} {GEANT4}),\
  }\href@noop {} {\bibfield  {journal} {\bibinfo  {journal} {Nucl. Instrum.
  Meth. A}\ }\textbf {\bibinfo {volume} {506}},\ \bibinfo {pages} {250}
  (\bibinfo {year} {2003})}\BibitemShut {NoStop}%
\bibitem [{\citenamefont {Frixione}\ and\ \citenamefont
  {Webber}(2002)}]{mcnlo}%
  \BibitemOpen
  \bibfield  {author} {\bibinfo {author} {\bibfnamefont {S.}~\bibnamefont
  {Frixione}}\ and\ \bibinfo {author} {\bibfnamefont {B.~R.}\ \bibnamefont
  {Webber}},\ }\href@noop {} {\bibfield  {journal} {\bibinfo  {journal} {JHEP}\
  }\textbf {\bibinfo {volume} {0206}},\ \bibinfo {pages} {029} (\bibinfo {year}
  {2002})},\ \Eprint {http://arxiv.org/abs/hep-ph/0204244}{arXiv:hep-ph/0204244
  [hep-ph]}\BibitemShut {NoStop}%
\bibitem [{\citenamefont {Corcella}\ \emph {et~al.}(2001)\citenamefont
  {Corcella} \emph {et~al.}}]{herwig}%
  \BibitemOpen
  \bibfield  {author} {\bibinfo {author} {\bibfnamefont {G.}~\bibnamefont
  {Corcella}} \emph {et~al.},\ }\href@noop {} {\bibfield  {journal} {\bibinfo
  {journal} {JHEP}\ }\textbf {\bibinfo {volume} {01}},\ \bibinfo {pages} {010}
  (\bibinfo {year} {2001})}\BibitemShut {NoStop}%
\bibitem [{\citenamefont {Binoth}\ \emph {et~al.}(2006)\citenamefont {Binoth},
  \citenamefont {Ciccolini}, \citenamefont {Kauer},\ and\ \citenamefont
  {Kramer}}]{gg2ww}%
  \BibitemOpen
  \bibfield  {author} {\bibinfo {author} {\bibfnamefont {T.}~\bibnamefont
  {Binoth}}, \bibinfo {author} {\bibfnamefont {M.}~\bibnamefont {Ciccolini}},
  \bibinfo {author} {\bibfnamefont {N.}~\bibnamefont {Kauer}}, \ and\ \bibinfo
  {author} {\bibfnamefont {M.}~\bibnamefont {Kramer}},\ }\href@noop {}
  {\bibfield  {journal} {\bibinfo  {journal} {JHEP}\ }\textbf {\bibinfo
  {volume} {12}},\ \bibinfo {pages} {046} (\bibinfo {year} {2006})},\ \Eprint
  {http://arxiv.org/abs/hep-ph/0611170}{arXiv:hep-ph/0611170}\BibitemShut
  {NoStop}%
\bibitem [{\citenamefont {Mangano}\ \emph {et~al.}()\citenamefont {Mangano}
  \emph {et~al.}}]{alpgen}%
  \BibitemOpen
  \bibfield  {author} {\bibinfo {author} {\bibfnamefont {M.}~\bibnamefont
  {Mangano}} \emph {et~al.},\ }\href@noop {} {\bibinfo  {journal} {JHEP {\bf
  07} (2003) 001}\ }\BibitemShut {NoStop}%
\bibitem [{\citenamefont {Campbell}\ \emph {et~al.}(2011)\citenamefont
  {Campbell}, \citenamefont {Ellis},\ and\ \citenamefont
  {Williams}}]{Campbell:2011bn}%
  \BibitemOpen
\bibfield  {journal} {  }\bibfield  {author} {\bibinfo {author} {\bibfnamefont
  {J.~M.}\ \bibnamefont {Campbell}}, \bibinfo {author} {\bibfnamefont {R.~K.}\
  \bibnamefont {Ellis}}, \ and\ \bibinfo {author} {\bibfnamefont
  {C.}~\bibnamefont {Williams}},\ }\href@noop {} {\bibfield  {journal}
  {\bibinfo  {journal} {JHEP}\ }\textbf {\bibinfo {volume} {07}},\ \bibinfo
  {pages} {018} (\bibinfo {year} {2011})},\ \Eprint
  {http://arxiv.org/abs/1105.0020}{arXiv:1105.0020 [hep-ph]}\BibitemShut
  {NoStop}%
\bibitem [{\citenamefont {Gavin}\ \emph {et~al.}(2011)\citenamefont {Gavin},
  \citenamefont {Li}, \citenamefont {Petriello},\ and\ \citenamefont
  {Quackenbush}}]{Gavin:2010az}%
  \BibitemOpen
  \bibfield  {author} {\bibinfo {author} {\bibfnamefont {R.}~\bibnamefont
  {Gavin}}, \bibinfo {author} {\bibfnamefont {Y.}~\bibnamefont {Li}}, \bibinfo
  {author} {\bibfnamefont {F.}~\bibnamefont {Petriello}}, \ and\ \bibinfo
  {author} {\bibfnamefont {S.}~\bibnamefont {Quackenbush}},\ }\href@noop {}
  {\bibfield  {journal} {\bibinfo  {journal} {Comput. Phys. Commun.}\ }\textbf
  {\bibinfo {volume} {182}},\ \bibinfo {pages} {2388} (\bibinfo {year}
  {2011})},\ \Eprint {http://arxiv.org/abs/1011.3540}{arXiv:1011.3540
  [hep-ph]}\BibitemShut {NoStop}%
\bibitem [{\citenamefont {Kersevan}\ and\ \citenamefont
  {Richter-Was}(2004)}]{acermc}%
  \BibitemOpen
  \bibfield  {author} {\bibinfo {author} {\bibfnamefont {B.~P.}\ \bibnamefont
  {Kersevan}}\ and\ \bibinfo {author} {\bibfnamefont {E.}~\bibnamefont
  {Richter-Was}},\ }\href@noop {} {\  (\bibinfo {year} {2004})},\ \Eprint
  {http://arxiv.org/abs/hep-ph/0405247}{arXiv:hep-ph/0405247
  [hep-ph]}\BibitemShut {NoStop}%
\bibitem [{\citenamefont {Sjostrand}\ \emph {et~al.}(2006)\citenamefont
  {Sjostrand}, \citenamefont {Mrenna},\ and\ \citenamefont {Skands}}]{pythia}%
  \BibitemOpen
  \bibfield  {author} {\bibinfo {author} {\bibfnamefont {T.}~\bibnamefont
  {Sjostrand}}, \bibinfo {author} {\bibfnamefont {S.}~\bibnamefont {Mrenna}}, \
  and\ \bibinfo {author} {\bibfnamefont {P.~Z.}\ \bibnamefont {Skands}},\
  }\href@noop {} {\bibfield  {journal} {\bibinfo  {journal} {JHEP}\ }\textbf
  {\bibinfo {volume} {0605}},\ \bibinfo {pages} {026} (\bibinfo {year}
  {2006})},\ \Eprint {http://arxiv.org/abs/hep-ph/0603175}{arXiv:hep-ph/0603175
  [hep-ph]}\BibitemShut {NoStop}%
\bibitem [{\citenamefont {Sherstnev}\ and\ \citenamefont
  {Thorne}(2008)}]{Sherstnev:2008dm}%
  \BibitemOpen
  \bibfield  {author} {\bibinfo {author} {\bibfnamefont {A.}~\bibnamefont
  {Sherstnev}}\ and\ \bibinfo {author} {\bibfnamefont {R.}~\bibnamefont
  {Thorne}},\ }\href@noop {} {\  (\bibinfo {year} {2008})},\ \Eprint
  {http://arxiv.org/abs/0807.2132}{arXiv:0807.2132 [hep-ph]}\BibitemShut
  {NoStop}%
\bibitem [{\citenamefont {Thorne}\ \emph {et~al.}(2009)\citenamefont {Thorne},
  \citenamefont {Martin}, \citenamefont {Stirling},\ and\ \citenamefont
  {Watt}}]{Thorne:2009ky}%
  \BibitemOpen
  \bibfield  {author} {\bibinfo {author} {\bibfnamefont {R.}~\bibnamefont
  {Thorne}}, \bibinfo {author} {\bibfnamefont {A.}~\bibnamefont {Martin}},
  \bibinfo {author} {\bibfnamefont {W.}~\bibnamefont {Stirling}}, \ and\
  \bibinfo {author} {\bibfnamefont {G.}~\bibnamefont {Watt}},\ }\href@noop {}
  {\  (\bibinfo {year} {2009})},\ \Eprint
  {http://arxiv.org/abs/0907.2387}{arXiv:0907.2387 [hep-ph]}\BibitemShut
  {NoStop}%
\bibitem [{\citenamefont {Pukhov}\ \emph {et~al.}(1999)\citenamefont {Pukhov},
  \citenamefont {Boos}, \citenamefont {Dubinin}, \citenamefont {Edneral},
  \citenamefont {Ilyin} \emph {et~al.}}]{calchep}%
  \BibitemOpen
  \bibfield  {author} {\bibinfo {author} {\bibfnamefont {A.}~\bibnamefont
  {Pukhov}}, \bibinfo {author} {\bibfnamefont {E.}~\bibnamefont {Boos}},
  \bibinfo {author} {\bibfnamefont {M.}~\bibnamefont {Dubinin}}, \bibinfo
  {author} {\bibfnamefont {V.}~\bibnamefont {Edneral}}, \bibinfo {author}
  {\bibfnamefont {V.}~\bibnamefont {Ilyin}},  \emph {et~al.},\ }\href@noop {}
  {\  (\bibinfo {year} {1999})},\ \Eprint
  {http://arxiv.org/abs/hep-ph/9908288}{arXiv:hep-ph/9908288
  [hep-ph]}\BibitemShut {NoStop}%
\bibitem [{\citenamefont {Pumplin}\ \emph {et~al.}(2002)\citenamefont
  {Pumplin}, \citenamefont {Stump}, \citenamefont {Huston}, \citenamefont
  {Lai}, \citenamefont {Nadolsky} \emph {et~al.}}]{cteq6}%
  \BibitemOpen
  \bibfield  {author} {\bibinfo {author} {\bibfnamefont {J.}~\bibnamefont
  {Pumplin}}, \bibinfo {author} {\bibfnamefont {D.}~\bibnamefont {Stump}},
  \bibinfo {author} {\bibfnamefont {J.}~\bibnamefont {Huston}}, \bibinfo
  {author} {\bibfnamefont {H.}~\bibnamefont {Lai}}, \bibinfo {author}
  {\bibfnamefont {P.~M.}\ \bibnamefont {Nadolsky}},  \emph {et~al.},\
  }\href@noop {} {\bibfield  {journal} {\bibinfo  {journal} {JHEP}\ }\textbf
  {\bibinfo {volume} {0207}},\ \bibinfo {pages} {012} (\bibinfo {year}
  {2002})},\ \Eprint {http://arxiv.org/abs/hep-ph/0201195}{arXiv:hep-ph/0201195
  [hep-ph]}\BibitemShut {NoStop}%
\bibitem [{\citenamefont {{ATLAS
  Collaboration}}(2010{\natexlab{b}})}]{ATLASsim}%
  \BibitemOpen
  \bibfield  {author} {\bibinfo {author} {\bibnamefont {{ATLAS
  Collaboration}}},\ }\href@noop {} {\bibfield  {journal} {\bibinfo  {journal}
  {Eur. Phys. J.}\ }\textbf {\bibinfo {volume} {C70}},\ \bibinfo {pages} {823}
  (\bibinfo {year} {2010}{\natexlab{b}})},\ \Eprint
  {http://arxiv.org/abs/1005.4568}{arXiv:1005.4568
  [physics.ins-det]}\BibitemShut {NoStop}%
\bibitem [{\citenamefont {{ATLAS
  Collaboration}}(2012{\natexlab{c}})}]{metPerf}%
  \BibitemOpen
  \bibfield  {author} {\bibinfo {author} {\bibnamefont {{ATLAS
  Collaboration}}},\ }\href@noop {} {\bibfield  {journal} {\bibinfo  {journal}
  {Eur. Phys. J.}\ }\textbf {\bibinfo {volume} {C72}},\ \bibinfo {pages} {1844}
  (\bibinfo {year} {2012}{\natexlab{c}})},\ \Eprint
  {http://arxiv.org/abs/1108.5602}{arXiv:1108.5602 [hep-ex]}\BibitemShut
  {NoStop}%
\bibitem [{\citenamefont {Moch}\ and\ \citenamefont {Uwer}(2008)}]{ttbar-xsec}%
  \BibitemOpen
  \bibfield  {author} {\bibinfo {author} {\bibfnamefont {S.}~\bibnamefont
  {Moch}}\ and\ \bibinfo {author} {\bibfnamefont {P.}~\bibnamefont {Uwer}},\
  }\href {\doibase 10.1103/PhysRevD.78.034003} {\bibfield  {journal} {\bibinfo
  {journal} {Phys. Rev.}\ }\textbf {\bibinfo {volume} {D78}},\ \bibinfo {pages}
  {034003} (\bibinfo {year} {2008})},\ \Eprint
  {http://arxiv.org/abs/0804.1476}{arXiv:0804.1476 [hep-ph]}\BibitemShut
  {NoStop}%
\bibitem [{\citenamefont {Kidonakis}(2011)}]{st-xsec-1}%
  \BibitemOpen
  \bibfield  {author} {\bibinfo {author} {\bibfnamefont {N.}~\bibnamefont
  {Kidonakis}},\ }\href {\doibase 10.1103/PhysRevD.83.091503} {\bibfield
  {journal} {\bibinfo  {journal} {Phys. Rev.}\ }\textbf {\bibinfo {volume}
  {D83}},\ \bibinfo {pages} {091503} (\bibinfo {year} {2011})},\ \Eprint
  {http://arxiv.org/abs/1103.2792}{arXiv:1103.2792 [hep-ph]}\BibitemShut
  {NoStop}%
\bibitem [{\citenamefont {Kidonakis}(2010)}]{st-xsec-2}%
  \BibitemOpen
  \bibfield  {author} {\bibinfo {author} {\bibfnamefont {N.}~\bibnamefont
  {Kidonakis}},\ }\href {\doibase 10.1103/PhysRevD.81.054028} {\bibfield
  {journal} {\bibinfo  {journal} {Phys. Rev.}\ }\textbf {\bibinfo {volume}
  {D81}},\ \bibinfo {pages} {054028} (\bibinfo {year} {2010})},\ \Eprint
  {http://arxiv.org/abs/1001.5034}{arXiv:1001.5034 [hep-ph]}\BibitemShut
  {NoStop}%
\bibitem [{\citenamefont {Ball}\ \emph {et~al.}(2010)\citenamefont {Ball},
  \citenamefont {Del~Debbio}, \citenamefont {Forte}, \citenamefont {Guffanti},
  \citenamefont {Latorre} \emph {et~al.}}]{Ball:2010de}%
  \BibitemOpen
  \bibfield  {author} {\bibinfo {author} {\bibfnamefont {R.~D.}\ \bibnamefont
  {Ball}}, \bibinfo {author} {\bibfnamefont {L.}~\bibnamefont {Del~Debbio}},
  \bibinfo {author} {\bibfnamefont {S.}~\bibnamefont {Forte}}, \bibinfo
  {author} {\bibfnamefont {A.}~\bibnamefont {Guffanti}}, \bibinfo {author}
  {\bibfnamefont {J.~I.}\ \bibnamefont {Latorre}},  \emph {et~al.},\
  }\href@noop {} {\bibfield  {journal} {\bibinfo  {journal} {Nucl. Phys.}\
  }\textbf {\bibinfo {volume} {B838}},\ \bibinfo {pages} {136} (\bibinfo {year}
  {2010})},\ \Eprint {http://arxiv.org/abs/1002.4407}{arXiv:1002.4407
  [hep-ph]}\BibitemShut {NoStop}%
\bibitem [{\citenamefont {Lai}\ \emph {et~al.}(2010)\citenamefont {Lai},
  \citenamefont {Huston}, \citenamefont {Mrenna}, \citenamefont {Nadolsky},
  \citenamefont {Stump} \emph {et~al.}}]{CT09MC2}%
  \BibitemOpen
  \bibfield  {author} {\bibinfo {author} {\bibfnamefont {H.-L.}\ \bibnamefont
  {Lai}}, \bibinfo {author} {\bibfnamefont {J.}~\bibnamefont {Huston}},
  \bibinfo {author} {\bibfnamefont {S.}~\bibnamefont {Mrenna}}, \bibinfo
  {author} {\bibfnamefont {P.}~\bibnamefont {Nadolsky}}, \bibinfo {author}
  {\bibfnamefont {D.}~\bibnamefont {Stump}},  \emph {et~al.},\ }\href@noop {}
  {\bibfield  {journal} {\bibinfo  {journal} {JHEP}\ }\textbf {\bibinfo
  {volume} {1004}},\ \bibinfo {pages} {035} (\bibinfo {year} {2010})},\ \Eprint
  {http://arxiv.org/abs/0910.4183}{arXiv:0910.4183 [hep-ph]}\BibitemShut
  {NoStop}%
\bibitem [{\citenamefont {Whalley}\ \emph {et~al.}(2005)\citenamefont
  {Whalley}, \citenamefont {Bourilkov},\ and\ \citenamefont
  {Group}}]{Whalley:2005nh}%
  \BibitemOpen
  \bibfield  {author} {\bibinfo {author} {\bibfnamefont {M.}~\bibnamefont
  {Whalley}}, \bibinfo {author} {\bibfnamefont {D.}~\bibnamefont {Bourilkov}},
  \ and\ \bibinfo {author} {\bibfnamefont {R.}~\bibnamefont {Group}},\
  }\href@noop {} {\  (\bibinfo {year} {2005})},\ \Eprint
  {http://arxiv.org/abs/hep-ph/0508110}{arXiv:hep-ph/0508110
  [hep-ph]}\BibitemShut {NoStop}%
\bibitem [{\citenamefont {Junk}(1999)}]{Junk:1999kv}%
  \BibitemOpen
  \bibfield  {author} {\bibinfo {author} {\bibfnamefont {T.}~\bibnamefont
  {Junk}},\ }\href@noop {} {\bibfield  {journal} {\bibinfo  {journal} {Nucl.
  Instrum. Meth. A}\ }\textbf {\bibinfo {volume} {434}},\ \bibinfo {pages}
  {435} (\bibinfo {year} {1999})},\ \Eprint
  {http://arxiv.org/abs/hep-ex/9902006}{arXiv:hep-ex/9902006
  [hep-ex]}\BibitemShut {NoStop}%
\bibitem [{\citenamefont {Cowan}\ \emph {et~al.}(2011)\citenamefont {Cowan},
  \citenamefont {Cranmer}, \citenamefont {Gross},\ and\ \citenamefont
  {Vitells}}]{Cowan:2010js}%
  \BibitemOpen
  \bibfield  {author} {\bibinfo {author} {\bibfnamefont {G.}~\bibnamefont
  {Cowan}}, \bibinfo {author} {\bibfnamefont {K.}~\bibnamefont {Cranmer}},
  \bibinfo {author} {\bibfnamefont {E.}~\bibnamefont {Gross}}, \ and\ \bibinfo
  {author} {\bibfnamefont {O.}~\bibnamefont {Vitells}},\ }\href@noop {}
  {\bibfield  {journal} {\bibinfo  {journal} {Eur. Phys. J.}\ }\textbf
  {\bibinfo {volume} {C71}},\ \bibinfo {pages} {1554} (\bibinfo {year}
  {2011})},\ \Eprint {http://arxiv.org/abs/1007.1727}{arXiv:1007.1727
  [physics.data-an]}\BibitemShut {NoStop}%
\end{thebibliography}%

\clearpage
\begin{flushleft}
{\Large The ATLAS Collaboration}

\bigskip

G.~Aad$^{\rm 48}$,
T.~Abajyan$^{\rm 21}$,
B.~Abbott$^{\rm 111}$,
J.~Abdallah$^{\rm 12}$,
S.~Abdel~Khalek$^{\rm 115}$,
A.A.~Abdelalim$^{\rm 49}$,
O.~Abdinov$^{\rm 11}$,
R.~Aben$^{\rm 105}$,
B.~Abi$^{\rm 112}$,
M.~Abolins$^{\rm 88}$
% ...
\bigskip
\\
% ...
$^{11}$ Institute of Physics, Azerbaijan Academy of Sciences, Baku, Azerbaijan\\
$^{12}$ Institut de F{\'\i}sica d'Altes Energies and Departament de F{\'\i}sica de la Universitat Aut{\`o}noma de Barcelona and ICREA, Barcelona, Spain\\
% ...
$^{21}$ Physikalisches Institut, University of Bonn, Bonn, Germany\\
% ...
$^{48}$ Fakult{\"a}t f{\"u}r Mathematik und Physik, Albert-Ludwigs-Universit{\"a}t, Freiburg, Germany\\
$^{49}$ Section de Physique, Universit{\'e} de Gen{\`e}ve, Geneva, Switzerland\\
% ...
$^{88}$ Department of Physics and Astronomy, Michigan State University, East Lansing MI, United States of America\\
% ...
$^{105}$ Nikhef National Institute for Subatomic Physics and University of Amsterdam, Amsterdam, Netherlands\\
% ...
$^{111}$ Homer L. Dodge Department of Physics and Astronomy, University of Oklahoma, Norman OK, United States of America\\
$^{112}$ Department of Physics, Oklahoma State University, Stillwater OK, United States of America\\
% ...
$^{115}$ LAL, Universit{\'e} Paris-Sud and CNRS/IN2P3, Orsay, France\\
% ...
\end{flushleft}


\end{document}